\documentclass{article}
\usepackage{arxiv}

\usepackage[utf8]{inputenc} 
\usepackage[T1]{fontenc}    
\usepackage{hyperref}       
\usepackage{url}            
\usepackage{booktabs}       
\usepackage{amsfonts}       
\usepackage{nicefrac}       
\usepackage{microtype}      
\usepackage{lipsum}
\usepackage{graphics}
\usepackage{graphicx}
\usepackage[labelformat=empty]{caption}
\usepackage{wrapfig}
\usepackage{float}
\usepackage[normalem]{ulem}
\usepackage{amsmath}
\usepackage{threeparttable}  
\usepackage{booktabs}
\usepackage{multirow}
\usepackage{bm}
\usepackage{authblk}
\usepackage{siunitx}
\newcommand{\corraddress}[1]{\def\@corraddress{#1}}
\newcommand{\corremail}[1]{\def\@corremail{#1}}
\newcommand{\authfn}[1]{}

\title{Quantitative Susceptibility Inversion Through \\ Parcellated Multiresolution Neural Networks \\ and K-Space Substitution}

\author[1 2 *]{Juan Liu}
\author[1 2 3]{Andrew S. Nencka}
\author[1 2 4]{L. Tugan Muftuler}
\author[1 4]{Brad Swearingen}
\author[1 3]{Robin Karr}
\author[1 2 3 *+]{Kevin M. Koch}

\affil[1]{Center for Imaging Research, Medical College of Wisconsin, Milwaukee, Wisconsin, 53226, USA}
\affil[2]{Biomedical Engineering, Marquette and Medical College of Wisconsin, Milwaukee, Wisconsin, 53226, USA}
\affil[3]{Radiology, Medical College of Wisconsin, Milwaukee, Wisconsin, 53226, USA}
\affil[4]{Neurosurgery, Medical College of Wisconsin, Milwaukee, Wisconsin, 53226, USA}
\affil[*]{Equally contributing authors.}
\affil[+]{Corresponding author: Kevin M. Koch, kmkoch@mcw.edu}

\corraddress{Kevin M Koch PhD, Department of Radiology, Medical College of Wisconsin, Milwaukee, Wisconsin, 53226, USA}
\corremail{kmkoch@mcw.edu}

\begin{document}
\maketitle

\begin{abstract}
\emph{Purpose:} Quantitative Susceptibility Mapping (QSM) reconstruction is a challenging inverse problem driven by poor conditioning of the field to susceptibility transformation.   State-of-art QSM reconstruction methods either suffer from image artifacts or long computation times, which limits QSM clinical translation efforts. To overcome these limitations, a deep-learning-based approach is proposed and demonstrated.

\emph{Methods:} An encoder-decoder neural network was trained to infer susceptibility maps on volume parcellated regions. The training data consisted of fabricated susceptibility distributions modeled to mimic the spatial frequency patterns of in-vivo brain susceptibility distributions.  Inferred volume parcels were recombined to form composite QSM.  This approach is denoted as ASPEN, standing for Approximated Susceptibility through Parcellated Encoder-decoder Networks.  ASPEN performance was evaluated relative to several well-established approaches on a gold-standard challenge dataset and on cohort of 200 study subjects.

\emph{Results:} ASPEN provided similar levels of quantitative accuracy compared to the evaluated established approaches on the gold standard ISMRM Challenge dataset, but qualitatively showed marked reductions in streaking artifacts and map blurring.  On the large-cohort dataset, ASPEN achieved the highest score compared with other methods in a multi-rater evaluation of streaking artifacts and map resolution.

\emph{Conclusion:} The proposed ASPEN approach can robustly infer susceptibility maps in near real-time on routine computational hardware.   This preliminary study establishes ASPEN's parity with existing approaches for quantitative accuracy on a well-curated gold standard dataset and further demonstrates its robustness to streaking artifacts across a large cohort of subjects.

\end{abstract}

\section{Introduction}
Quantitative susceptibility mapping (QSM) is a magnetic resonance imaging (MRI) technique that uses Larmor frequency off-resonance  measurements to estimate bulk material magnetic susceptibilities. QSM has been used to study iron content \cite{haacke2005imaging,deistung2013toward,bilgic2012mri,zheng2013measuring,langkammer2012quantitative}, blood products \cite{zhang2015quantitative}, myelin \cite{wisnieff2015quantitative}, and endogenous contrast \cite{xu2015quantification}. QSM has been shown to be useful in assessing neurodegeneration \cite{bilgic2012mri,acosta2013vivo, lotfipour2012high} and multiple sclerosis \cite{wisnieff2015quantitative,langkammer2013quantitative}. 
Other QSM applications have explored detecting calcification in brain tumors \cite{deistung2013quantitative,chen2014intracranial}, iron overload in the liver \cite{sharma2015quantitative}, the delineation of white matter tracts in the brain using susceptibility tensor imaging \cite{li2014mean}, and mild traumatic brain injury~\cite{koch2018quantitative}.  
 
 Susceptibility maps are generated by extracting Larmor frequency shifts from MRI data and solving for the source tissue susceptibility. To date, all QSM methods rely on a dipolar convolution that relates source susceptibility to induced Larmor frequency offsets ~\cite{salomir2003fast, marques2005application}. While the forward relationship of this model (source to field) can be efficiently computed using Fast-Fourier-Transforms (FFT), a k-space singularity in the applied convolution filter results in an ill-conditioned relationship in the inverse model (field to source). Though clinically challenging, acquiring data at multiple orientations to the magnetic field remains the empirical gold-standard for \emph{in vivo} QSM assessment, as the additional field data sufficiently improves the conditioning of the inversion algorithm~\cite{liu2009calculation}. In the absence of multiple-orientation acquisitions, single-orientation susceptibility maps are computed by either thresholding of the convolution operator~\cite{shmueli2009magnetic, wharton2010susceptibility, haacke2010susceptibility} or use of more sophisticated regularization methods \cite{de2008quantitative, de2010quantitative, liu2011morphology, bilgic2014fast}.  

In single-orientation QSM, streaking artifacts propagating through the entire volume are a common nuisance\cite{li2015method,wang2015quantitative}. In practice, these artifacts often are exacerbated by locally extreme susceptibility values (such as blood products) or erroneous Larmor frequency-offset estimates, specifically in regions near the boundaries of the brain. To reduce these artifacts, regional masking of the extreme frequency offset regions and increasing regularization constraints can be applied. However, increased masking erodes potentially useful imaging data along the brain boundaries.  In addition, higher regularization penalties reduce map resolutions and can influence the quantitative values of the resulting susceptibility map \cite{wharton2010whole}. Furthermore, these workflow and algorithmic modifications often require manual intervention, and in many cases still fail to eliminate artifacts.

Recently, deep neural networks have emerged as promising alternatives to iterative methods for inverse problems such as denoising \cite{vincent2010stacked, xie2012image}, image reconstruction \cite{sun2016deep}, and super-resolution \cite{dong2014learning}. 
Deep neural networks are advantageous for solving inverse problems because of their representation learning approach. Different layers of the network learn different representations of input data, which allows for learning the generic set of the inverse functions.

Initial preliminary investigations have already shown the potential of deep neural networks for QSM inversion. In particular, the recently described DeepQSM approach \cite{rasmussen2018deepqsm} utilized a modified 3D U-Net neural network \cite{ronneberger2015u} and synthetic 3D images simulated using basic geometric shapes as training dataset to perform QSM inversion.   

Another recently described approach, known as QSMnet\cite{yoon2018quantitative}, utilized multiple head orientation phase maps as the training input and Calculation Of Susceptibility through Multiple Orientation Sampling (COSMOS) maps \cite{liu2009calculation} as the training label. An initial analysis of this method using five different head orientations scans from one subject showed that QSMnet can produce high fidelity QSM maps comparable to COSMOS method, irrespective of the input head orientations. 

Finally, another deep neural network proposed by Gong \cite{gong2018qsm} utilized multiple inputs to the neural network model: 1) magnitude combined echo images, 2) raw tissue phase using Laplacian Boundary Value (LBV)\cite{zhou2014background} background removal, 3) TKD estimated QSM and 4) closed-form L$_2$ regularized QSM estimation \cite{bilgic2014fast}. The model utilized a 2D encoder-decoder network architecture, and was trained with COSMOS multi-orientation derived QSM as the output label. 

Here, an alternative QSM deep neural network using encoder-decoder architecture is described and analyzed. This QSM inversion approach relies upon three fundamental concepts: 1) multi-resolution spatially parcellated inversion to account for near and distant dipolar interactions using encoder-decoder network architecture, 2) neural-network inversion trained on fully simulated data generated from spatial harmonic analysis of acquired data, and 3) convolution k-space segmentation and replacement to maximize the utility of the trained neural network. 

The presented QSM inversion approach is denoted as ASPEN, which stands for Approximated Susceptibility through Parcellated Encoder-decoder Networks.  

In this preliminary study, the ASPEN inversion approach was applied to the 2016 ISMRM QSM challenge dataset~\cite{langkammer2018quantitative} and compared against several commonly utilized inversion approaches, including Truncated K-Space Division (TKD) inversion ~\cite{shmueli2009magnetic}, closed-form L$_2$ regularized inversion (CFL2)  ~\cite{bilgic2014fast}, and Morphology Enabled Dipole Inversion (MEDI) ~\cite{liu2012morphology}. 

In addition, the ASPEN  was applied to 200 datasets collected within a study of sports concussion using a clinically viable QSM acquisition protocol (4 minute acquisition time).  This provided a test of the inversion performance on a non-isotropic dataset utilizing subjects with imperfect motion compliance during data acquisition. 
Within this dataset, ASPEN was again compared with the aforementioned conventional approaches.  To investigate practical quality measures of the reconstructed maps, a multi-reader study was performed to assess streaking artifacts and image sharpness.

\section{Methods}
\subsection{ASPEN:  Training dataset}
\begin{figure}[h]
\begin{center}
\includegraphics[width=4.8in]{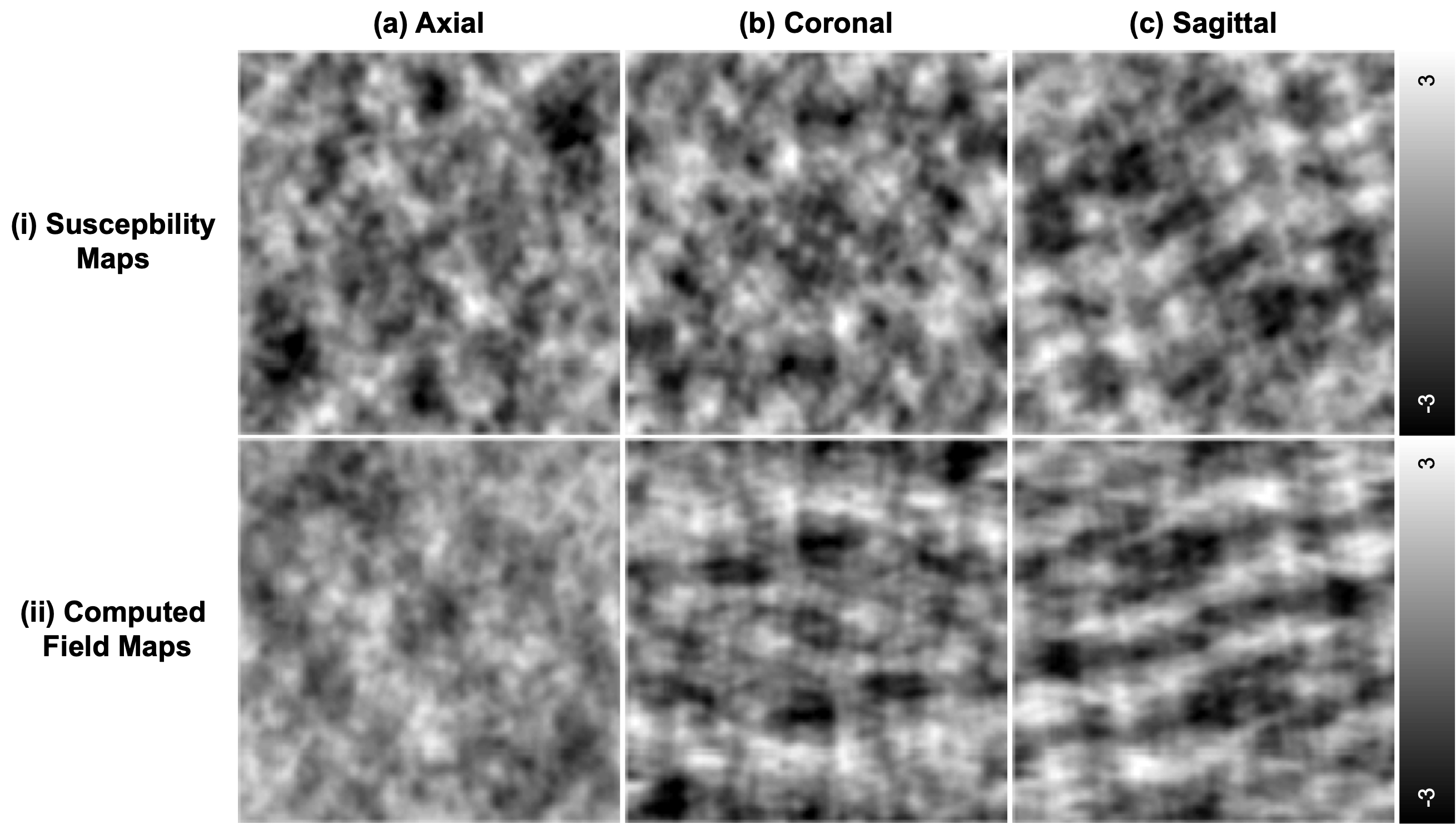}
\caption{\textbf{\small {Figure 1. One representative slice from a 3D volume of synthetic training data. The susceptibility maps and forward field maps have 128x128x128 voxels with isotropic 1.06 mm resolution. Both the susceptibility maps and forward field maps apply data normalization with mean zero and standard deviation one for training.}}}
\label{synQSMData} 
\end{center}
\end{figure}

Training data for ASPEN neural networks were constructed by randomly fabricating susceptibility distributions so as to mimic the spatial frequency patterns of \emph{in vivo} brain susceptibility estimates. These susceptibility distributions served as the label for the neural network training. Spatial frequency power spectra were estimated from QSM estimates using COSMOS (for ISMRM challenge dataset) or MEDI (n=50 for the sports concussion cohort).    
The simulated susceptibility maps were generated using inverse Fourier transform of the square root of the cohort-averaged spectral power density of the the target susceptibility with element-wise random phase swaps. The simulated susceptibility maps were then convolved with the dipole kernel to create input field perturbation maps to be used as training inputs for the neural network. Figure~\ref{synQSMData} provides sample maps from the synthesized training dataset. 



\subsection{ASPEN:  Network Design}
The ASPEN approach utilizes a 3D encoder-decoder deep neural network. The input to the network are the local field perturbation maps and the output is an estimate of the source magnetic susceptibility tensor at each input voxel. The applied encoder-decoder network architecture utilizes skip connections between the encoding path and decoding path, which can effectively transfer local feature information from the encoding path to the decoding path and facilitate faster training~\cite{long2015fully, ronneberger2015u}. 

\begin{figure}[h]
\begin{center}
\includegraphics[width=4.8in]{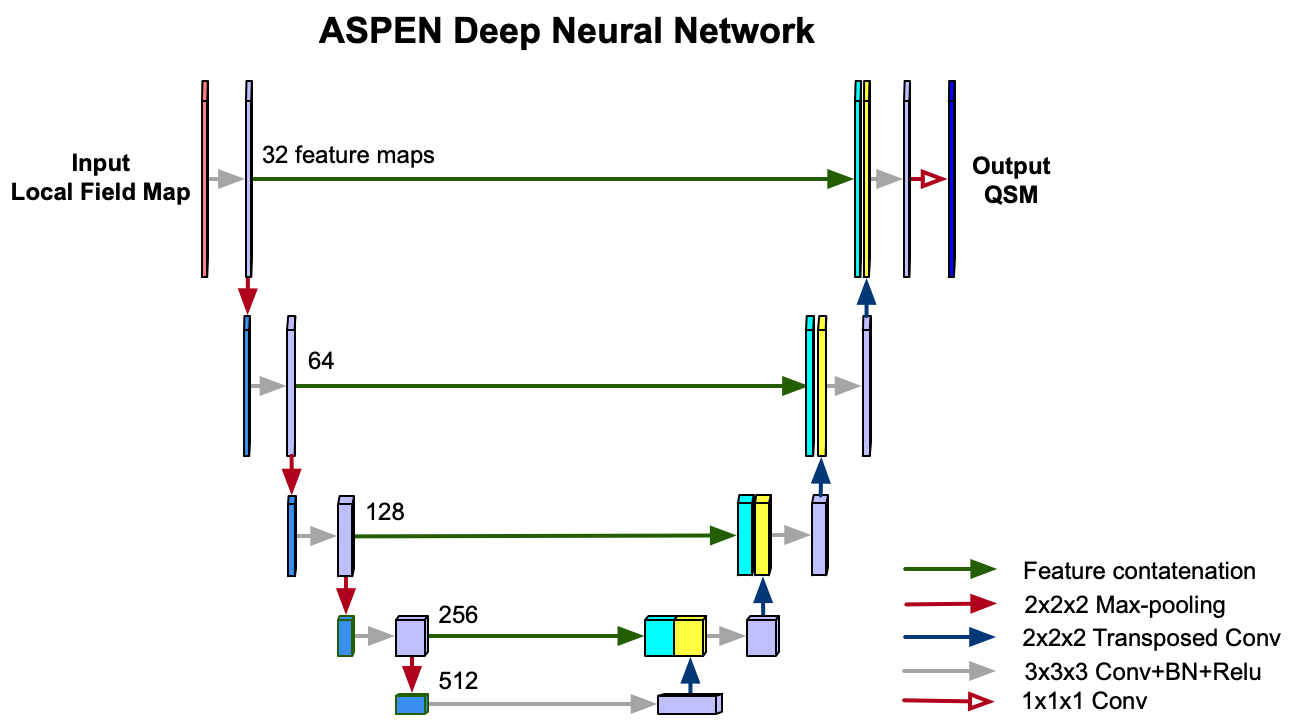}
\caption{\textbf{\small {Figure 2. Network structure of ASPEN. A 3D encoder-decoder neural network was designed with 9 convolutional layers with 3x3x3 kernels followed by batch normalization and ReLU as activation, 1 convolutional layer with kernel size 1x1x1 with linear activation, 4 max-pooling layers with strides 2x2x2, 4 transposed convolutional layers with 3x3x3 kernels and strides 2x2x2, and 4 feature concatenations.}}}
\label{ASPENNN}
\end{center}
\end{figure}

The ASPEN encoder-decoder network, as indicated in Figure~\ref{ASPENNN},  consists of 9 convolutional layers with 3x3x3 kernels, 9 batch normalization layers, 9 rectified linear unit (ReLU) layers,
4 max-pooling layers using stride 2x2x2, 4 transposed convolutional layers with kernel 3x3x3 and stride 2x2x2, and 4 feature contracting paths. The encoding path has four groups, with each group containing one convolutional layer with 3x3x3 kernels, batch normalization, ReLU activation, all connected by a max-pooling layer. The decoding path was connected via a transposed convolutional layer, with additional feature concatenation layers from the encoding layers. Finally, a 1x1x1 convolutional layer with linear activation was applied to output the susceptibility estimations. 

A L$_1$ loss function and ADAM optimizer~\cite{kingma2014adam} were used to train the aforementioned network. The initial learning rate was set as 0.001, with an exponential decay of the rate occurring at every 200 steps with a time constant of 0.9. Due to the deep nature of this neural network and the heavy computation required to train it, 10000 datasets were used for training each model with batch size 8. The neural network was trained and evaluated in Keras using Tensorflow as backend on four NVIDA Tesla K80 GPU.  

Due to the different target QSM map resolutions for each of the test scenarios, two ASPEN network models were trained. For ISMRM QSM challenge dataset with voxel size isotropic 1.06 mm, the ASPEN network was trained to independently invert 3D parcels of 128x128x128 voxels. For the sports concussion test cohort, with voxel size 0.5x0.5x2.0 mm$^3$, the ASPEN network was trained to invert 3D parcels of 192x192x64 voxels. In the prediction stage, segmentation of full-resolution input volumes into the parcels with parcel size equal to neural network input data size was performed using 32x32x32 overlap regions. After QSM inference using the trained encoder-decoder networks, the parcels were combined to form a composite image. In order to avoid edge artifacts in the combined ASPEN QSM maps, 8x8x8 voxel boundaries (within the 32x32x32 overlap region) were discarded before parcel combination. Linear regression utilizing the overlapping regions between parcels was used to adjust the relative scaling and bias between the individual parcels. Final QSM maps were then computed via k-space substitution. A hard threshold of 0.2 was utilized to define the region of k-space where the forward field map of QSM estimation was substituted into the k-space of input field map. After this substitution, inverse Fourier transformation yielded the final ASPEN QSM estimate.

\subsection{Performance Evaluation of ASPEN}
\subsubsection*{2016 QSM reconstruction Challenge Dataset} 
As previously presented by Langkammer et al~\cite{langkammer2018quantitative}, this data set was acquired from a healthy 30-year-old female subject.   Data for COSMOS and susceptibility-tensor computation was computed using a heavily accelerated wave-CAIPI~\cite{bilgic2015wave} acquisition at 1.06mm isotropic resolution collected at 12 different head orientations at a single echo time.  Conventional single-orientation QSM data was collected at the same isotropic resolution for 4 echo times.  Further details on these data acquisitions are available in Langkammer et al's review of the challenge results~\cite{langkammer2018quantitative}.   Qualitatively, this challenge dataset is of very high quality with minimal motion artifact.   


Using this gold-standard evaluation dataset, ASPEN performance was compared to TKD, CLF2, and MEDI approaches. TKD and CLF2 results were provided publically by the QSM challenge organizers. The MEDI solution was calculated with a manually chosen regularization parameter of 3000 yielding the best trade-off between quantification accuracy and artifacts.  All methods were evaluated against the "gold standard" susceptibility-tensor imaging (STI) (3,3) component computational result provided with the challenge dataset ~\cite{liu2010susceptibility}.  Estimation errors from each technique were computed using root mean squared error (RMSE), high-frequency error norm (HFEN), structural similarity (SSIM) index, and region-of-interest quantification errors that precisely follow the approach utilized in ISMRM Challenge analysis~\cite{langkammer2018quantitative}.   


\subsubsection*{Large Cohort Evaluation} 
Evaluation of ASPEN and the aforementioned QSM techniques on a typical study cohort was performed using 200 of QSM datasets acquired within a large local sports concussion study. The study was approved by the local institutional human research review board. Participants provided written consent, or assent and parental consent if minors were obtained. 
 The full analysis cohort consisted of injured and matched control subjects scanned longitudinally in 4 scheduled visits over a 45 day period.  Of the 971 collected QSM datasets collected in this study, 200 were randomly selected for use in the present analysis. 
 
MR imaging for the sports concussion cohort was performed on a 3T MRI scanner (GE Healthcare MR750) using a 32ch MRI head receive array.  QSM data from this study was collected by saving the raw k-space data from a commercially available susceptibility-weighted software application (SWAN, GE Healthcare). The data acquisition parameters were as follows: in-plane data matrix - 320x256, field of view - 24 cm, voxel size - 0.5x0.5x2.0 mm$^3$, echo spacing - 7 ms, 4 echo times - [10.4, 17.4, 24.4, 31.4] ms, repetition time - 58.6 ms, autocalibrated parallel imaging factors - 3x1, acquisition time - 4 min.

Complex multi-echo images were reconstructed from raw k-space data. The brain masks were obtained using the SPM tool~\cite{brett2002region}. After background field removal using the LBV method~\cite{zhou2014background}, susceptibility inversion was performed using the TKD, CFL2, MEDI, and ASPEN. The publicly available MEDI toolbox~\cite{MEDI_cornell_WEB} and closed-form L2 regularization code~\cite{bilgic2014fast} were utilized for MEDI and CFL2 QSM calculation. For TKD, the filter truncation value was set to 50 to trade-off the streaking artifacts and quantification accuracy. For MEDI and CFL2, the regularization factor was set to default values of 1000 and 0.029 respectively.

To compare the quality of susceptibility maps computed using each technique, the reconstructed maps were evaluated in axial, coronal, and sagittal planes. Three raters experienced with management and quality oversight of QSM data were trained to perform ranking of each technique of the level of streaking artifacts and sharpness of each map for each subject in the evaluation cohort. The computed QSM maps for each method were randomly displayed for each subject. For streaking artifacts evaluation, the rater ranked the four QSM maps from one to four, based on the severity and number and severity of streaks (4 being the best appearing map). For map sharpness, the maps were also ranked from one to four, with four being the sharpest (highest resolution) map.    Performance statistics were computed within and across the raters.   Inter-rater agreement was analyzed using intra-class correlation analysis and unilateral agreement counts.   

\section{RESULTS}
\begin{figure}[h]
\begin{center}
\includegraphics[width=5.8in]{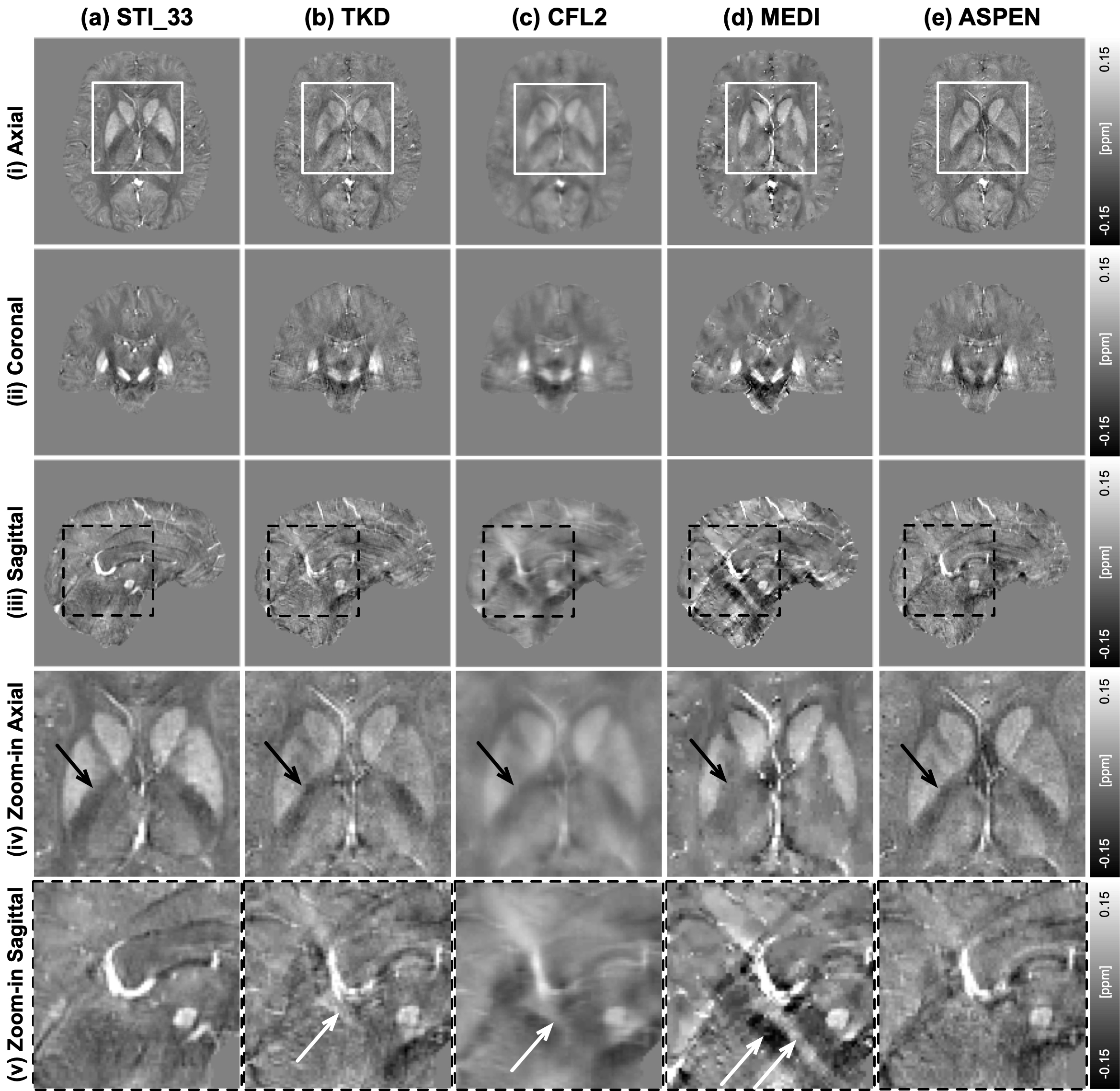}
\caption{\textbf{\small {Figure 3. STI (3,3), TKD, CFL2, MEDI, and ASPEN susceptibility maps from 2016 QSM reconstruction challenge in axial (i), coronal (ii), and sagittal (iii) views. The last two rows (iv,v) are zoomed maps, as indicated by boxes in (i,iii).  Solid black arrows indicate regions of noticeably improved performance of ASPEN in reproducing the STI (3,3) map estimates.   Streak artifacts indicated with white dashed arrows are not visible in the ASPEN maps (e).}}}
\label{qsmChallengeMaps}
\end{center}
\end{figure}

Figure~\ref{qsmChallengeMaps} illustrates susceptibility maps from the 2016 ISMRM QSM Challenge dataset reconstructed using ASPEN and TKD are displayed in 3 reformatted planes(rows i,ii,ii).   Streaking artifacts are clearly identified in the sagittal reformat for both CFL2 and MEDI maps.   In addition, the CFL2 map shows clearly compromised spatial resolution relative to the other maps.   These observations are amplified in the zoomed maps in the last two rows (iv, v), where clear ASPEN performance improvements in reproducing the fine structure of the STI (3,3) map are indicated by solid black arrows (white solid box in axial plane) and subtle streaking artifacts removed by the ASPEN approach are identified and indicated by the white arrows (black dashed box in sagittal plane).   

\begin{figure}[h!]
\begin{center}
\includegraphics[width=5.8in]{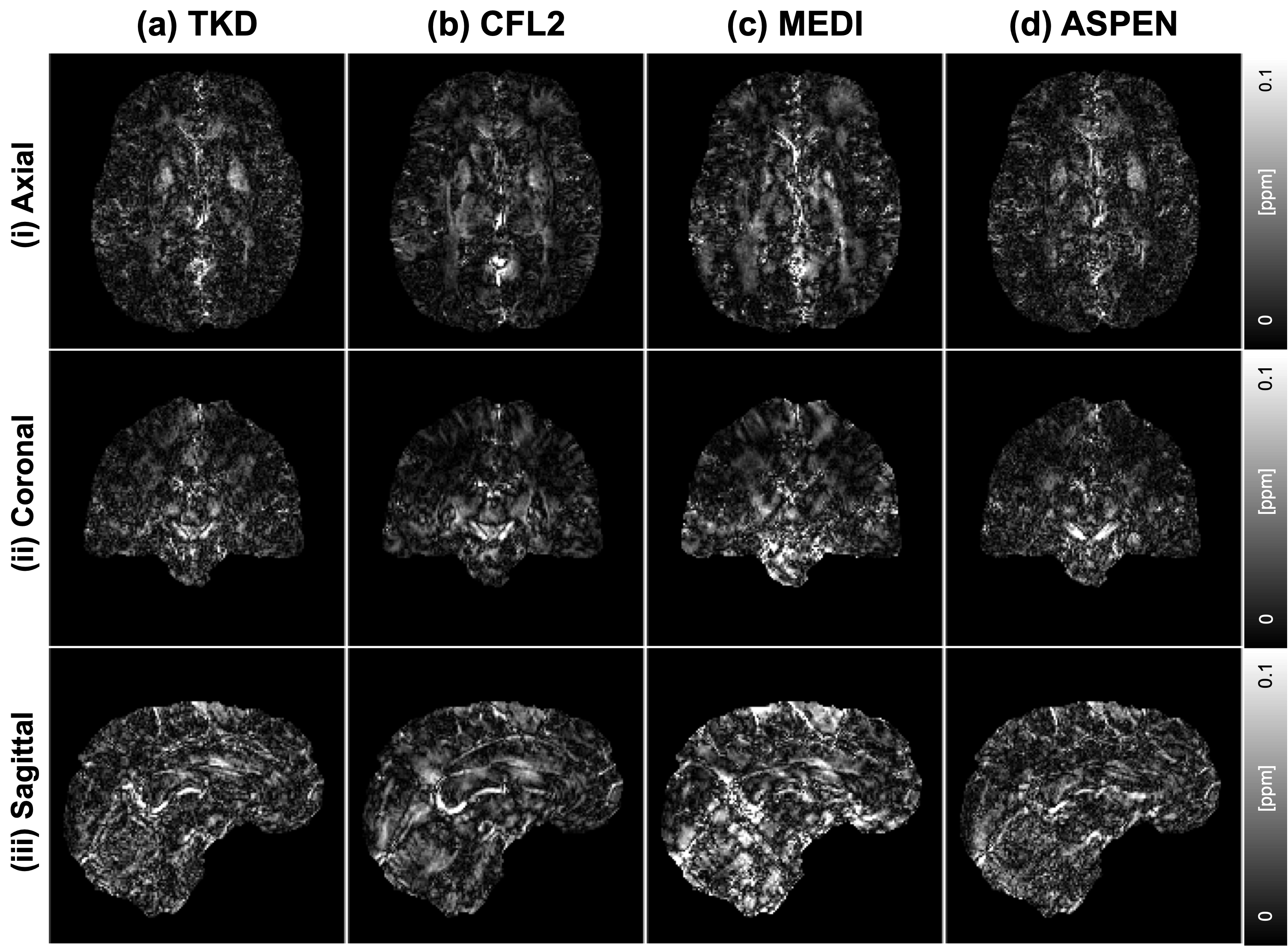}
\caption{\textbf{\small {Figure 4. Absolute error maps of TKD, CFL2, MEDI, and ASPEN susceptibility maps relative to the STI (3,3) estimate from 2016 QSM reconstruction challenge. }}}
\label{qsmChallengeErrors}
\end{center}
\end{figure}

Figure~\ref{qsmChallengeErrors} provides error maps of each evaluated technique in 3 reformatted planes (i,ii,iii) relative to the STI (3,3) estimate. Qualitatively, the errors of all techniques appear to be similar. Following the precedent set by the QSM Challenge dataset analysis~\cite{langkammer2018quantitative}, several performance metrics are reported in Table~\ref{qsmChallengeMapsMetrix}.  Quantitatively, the ASPEN approach shows similar accuracy performance relative to the evaluated techniques.

\begin{figure}[h]
\begin{center}
\includegraphics[width=5.8in]{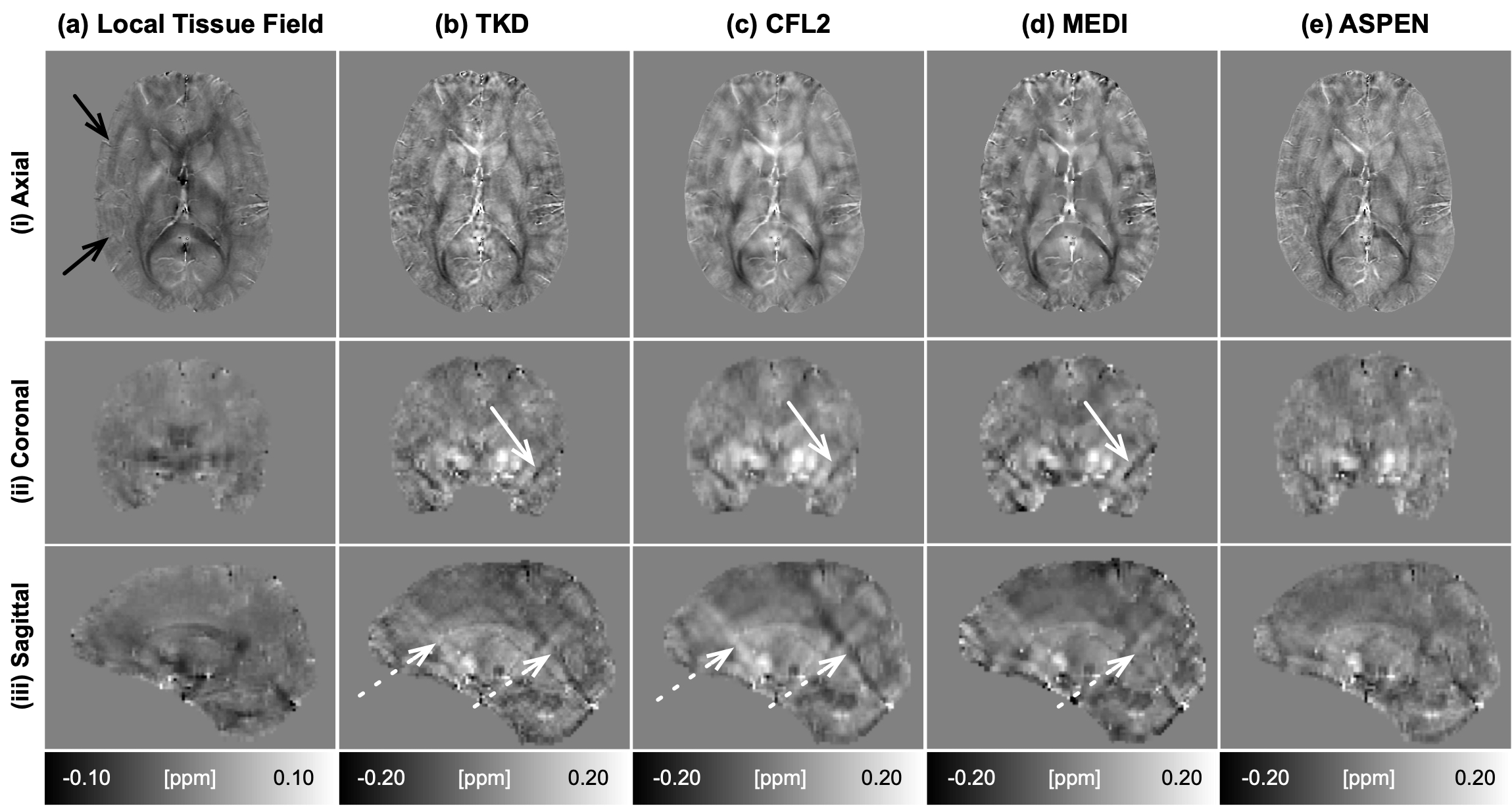}
\caption{\textbf{\small {Figure 5. Comparison of QSM reconstructed by TKD, CFL2, MEDI, and ASPEN on a representative dataset from the sports concussion cohort. Solid black arrows indicate subtle, but common, motion artifacts in tissue field that increased the difficulty of QSM inversion.  White arrows indicate streaking artifacts in the coronal (ii) and sagittal (iii) reformatted images of TKD, CFL2 and MEDI estimations.}}}
\label{srcQSMFigure1}
\end{center}
\end{figure}

Figure~\ref{srcQSMFigure1} shows QSM of the four evaluated techniques in axial (i), coronal (ii), and sagittal (iii) views on a representative dataset from the large cohort concussion study.
This subject showed some minor motion during the acquisition, which manifested as subtle artifacts in the input tissue field (black arrows in [i,a]) and added confounds to the QSM inversion.  Such motion is commonly encountered in clinical and symptomatic disease cohorts.

When inverting the tissue field of this subject, TKD preserves subtle microstructural features but suffers noticeable streaking artifacts in both the coronal (solid white arrows in [ii,b] ) and sagittal (dashed white arrows in [iii,b]) planes. The CFL2 maps show both blurring (i,ii,ii,c) and substantial streaking (arrows in [ii,iii,c]).  MEDI maps show some loss of resolution (i,ii,iii,a) and less streaking artifact (arrows in [ii,iii,d]). Qualitatively, ASPEN maps (e) show the best overall quality with least streaking artifacts and best depiction of tissue microstructure.

\begin{figure}[h]
\begin{center}
\includegraphics[width=5.8in]{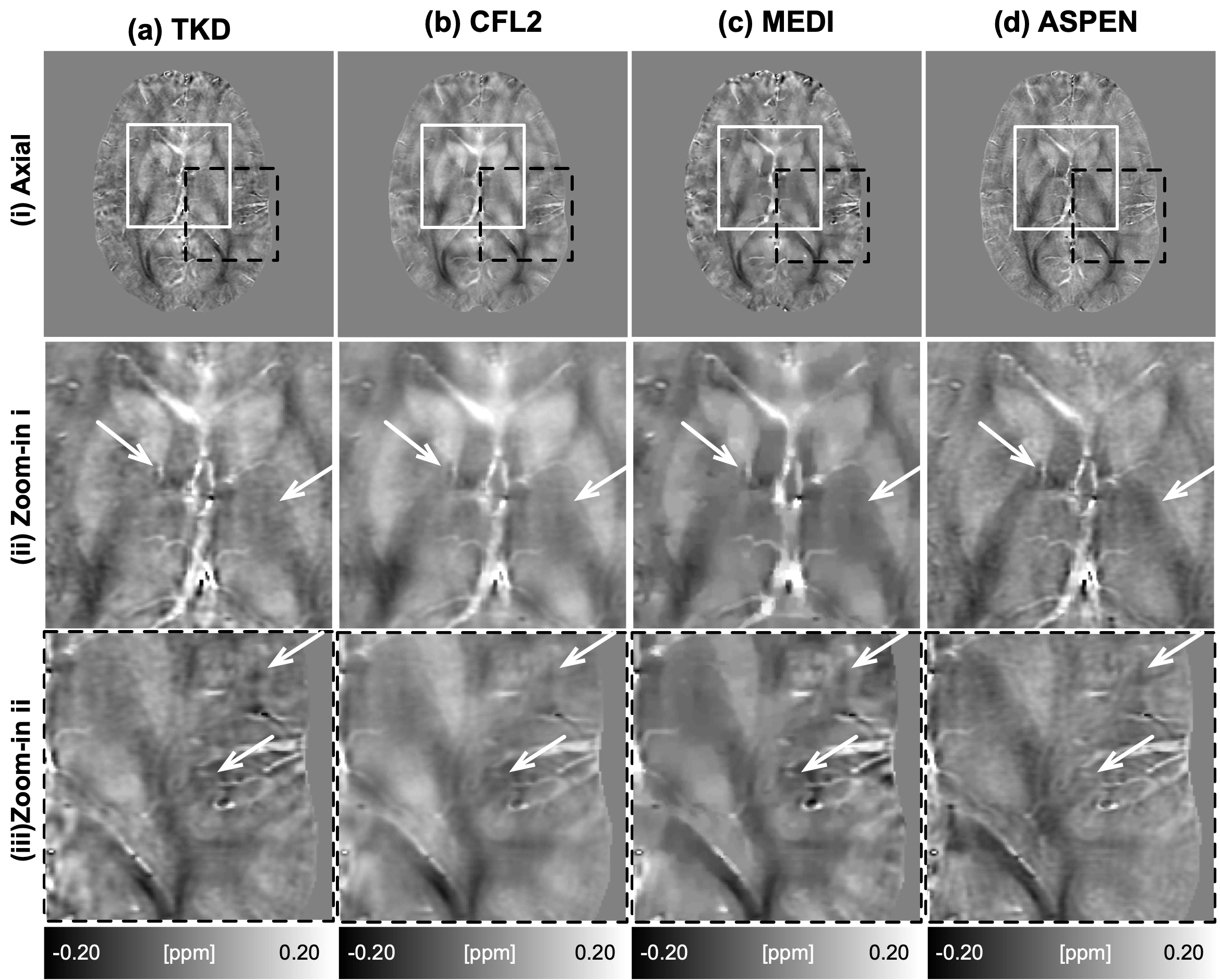}
\caption{\textbf{\small {Figure 6. Detailed visual comparison (axial view) and zoom-in visualization (ii,iii) of a representative concussion QSM study case. In TKD, CFL2 and MEDI images, substantial image blurring is clearly visible. ASPEN (d) maps have superior image sharpness and well-preserved details, as indicated by white arrows.}}}
\label{srcAxialFig}
\end{center}
\end{figure}

\begin{figure}[h]
\begin{center}
\includegraphics[width=5.8in]{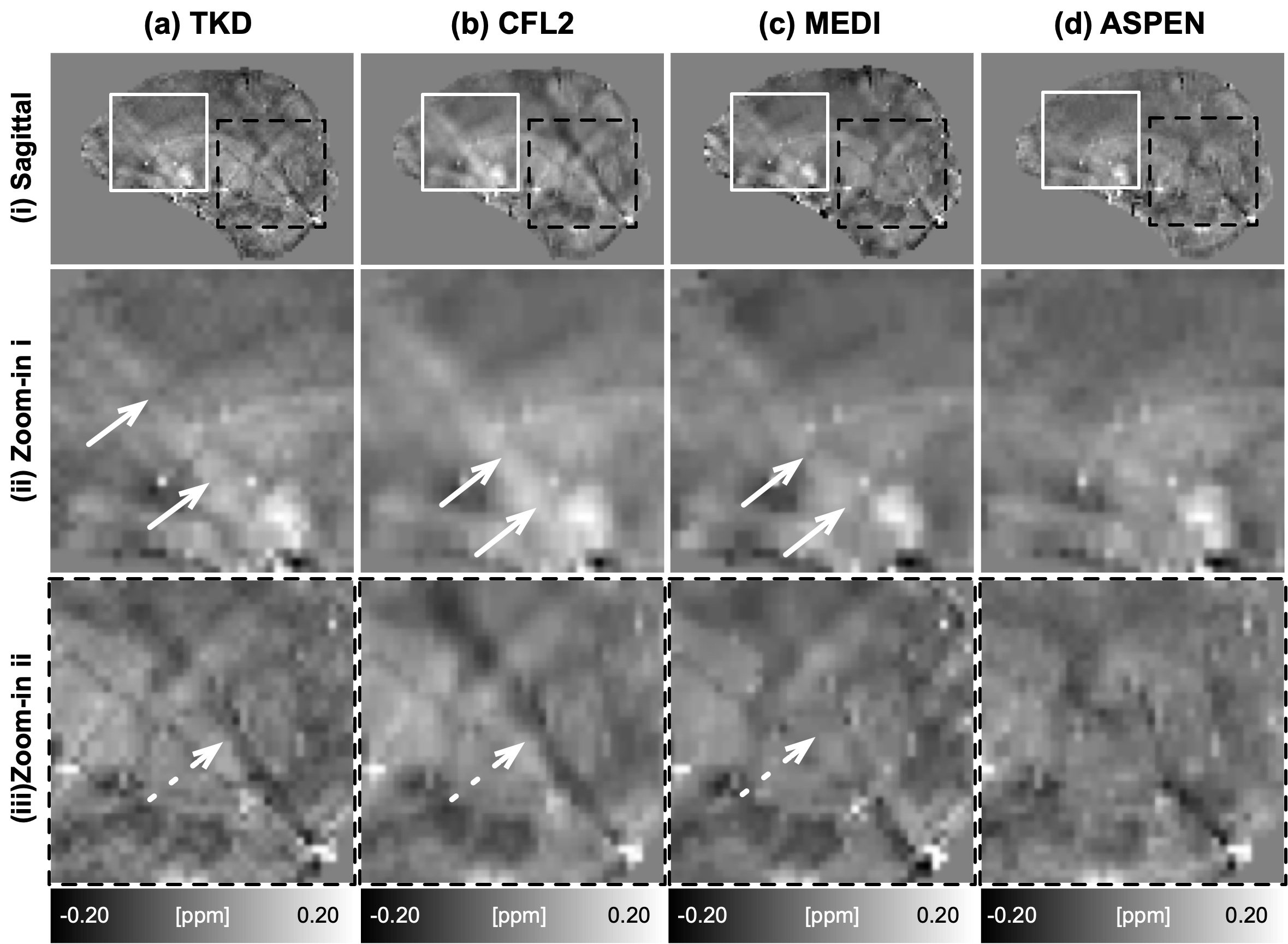}
\caption{\textbf{\small {Figure 7. Detailed visual comparison (sagittal view) and zoom-in visualization (ii,iii). In TKD, CFL2, and MEDI images, streaking artifacts are clearly visible in zoom-in images, as indicated by white arrows. The streaks are greatly reduced in the ASPEN maps (d).}}}
\label{srcSagittalFig}
\end{center}
\end{figure}

Figure~\ref{srcAxialFig} provides zoomed axial (i) views that clearly demonstrate the superior sharpness of the ASPEN (d) maps. Although TKD (a) also shows well-preserved microstructure, there is compromised integrity at tissue boundaries (white arrows). Again, the CFL2 (b) and MEDI (c) maps show increased blurring due to their heavy use of spatial regularization to reduce streaking artifacts. In Figure~\ref{srcSagittalFig}, the sagittal (ii,iii) zoomed views show that ASPEN (d) has minimal observed streaking artifacts, while other methods (a,b,d) have clearly visible streaking artifacts (white arrows).



\begin{figure}[h]
\begin{center}
\includegraphics[width=4.8in]{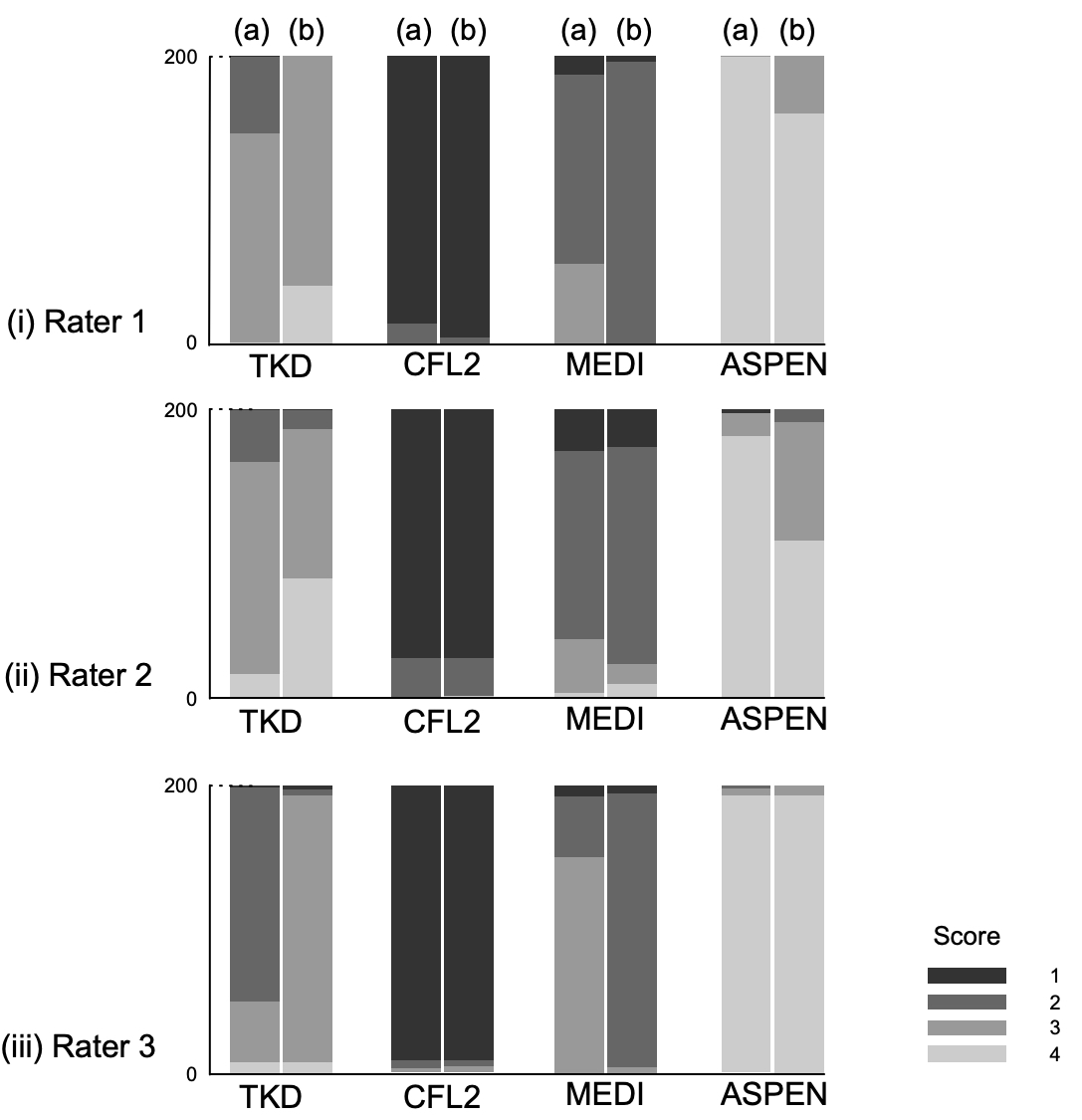}
\caption{\textbf{\small {Figure 8. Stack plot of the blinded visual rating score of four QSM inversion algorithms of three raters (i,ii,iii) on 200 QSM datasets. The first column (a) reflects streaking artifact score according to the indicated color map.   The second column (b) is map sharpness assessment result.  ASPEN provided consistently better scoring than the other approaches in both streaking and sharpness. }}}
\label{raterGraph}
\end{center}
\end{figure}

Figure~\ref{raterGraph} provides a bar graph summarizing the rater (i,ii,ii) scoring of the four QSM approaches for streaking (a) and sharpness (b) on the 200 datasets from the concussion cohort. Across all raters, ASPEN showed statistically significant high rating score relative to the other methods in both streaking artifacts and map resolution. Table~\ref{raterMetrix2} provide the statistical results of the rater study. Agreement between raters was strong, with respective mean ICCs of 0.81 and 0.86 for streaking and sharpness rankings. ASPEN was found to be the best performer in streaking reduction by all three reviewers in 80$\%$ of the test cases.   Conversely, CFL2 was universally rated to be the worst performer in both streaking and sharpness in roughly 80$\%$ of the cases.   This is in contrast to the seemingly strong quantitative performance of CFL2 on the pristine QSM Challenge dataset (Table~\ref{qsmChallengeMapsMetrix}).

\section{DISCUSSION}
This study has introduced ASPEN, a QSM inversion approach that utilizes volume parcellation, deep encoder-decoder neural networks, and k-space substitution. The preliminary performance results of this approach applied to the 2016 ISMRM QSM Challenge dataset and a large cohort of concussion study subjects show encouraging preliminary performance.  In a qualitative blinded rater study of 200 datasets, ASPEN was conclusively shown to have less streaking artifacts and improved image sharpness compared to established QSM approaches. 

The QSM Challenge dataset is a well-curated isotropic acquisition, and generally displays far better performance than non-isotropic datasets acquired on untrained study participants or clinical subjects.    As such, the streaking artifacts in conventional techniques displayed in Figures~\ref{qsmChallengeMaps} are very modest.   More substantial streaks are seen in the representative dataset collected on an athlete in the analyzed sports concussion cohort (Figures~\ref{srcQSMFigure1},~\ref{srcSagittalFig}).  This cohort consisted of subject datasets moderately compromised data quality, due to the scanning of injured and control athletes that exhibited typical levels of motion during signal acquisition.  A manual quality control evaluation of this dataset found the source QSM magnitude images to have mean score of 2.6/3. In this dataset, streaking artifacts substantially reduced using the ASPEN approach, which can be attributed to the lack of reliance on information from mathematically compromised regions of the inverted dipole kernel.   From a broad perspective, ASPEN is utilizing deep learning to estimate the dipolar convolution kernel in these compromised regions.   As demonstrated by the results of this preliminary study, this approach can produce QSM with negligible streaks, high resolution, and generalized quantitative accuracy that is comparable to existing approaches.   

The results of this preliminary study provide further feasibility confirmation for the utility of deep neural networks in performing QSM inversion. Of crucial importance for practical QSM clinical application, a trained ASPEN network can perform QSM inversion in a matter of seconds on standard reconstruction computers. In addition, this approach can be trained to deal with confounding effects in practical QSM application, such substantial hemorrhages in stroke or traumatic brain injury patients.


Compared to other recently published QSM deep learning methods, ASPEN has several advantages. First, ASPEN uses spatial power spectra to generate training datasets which mimic the spatial frequency patterns of in-vivo brain susceptibility. This approach is an easily transferable method to train networks for QSM acquisitions of different resolutions and specific applications. Similar to the approach utilized for DeepQSM \cite{rasmussen2018deepqsm}, but different from the approaches in QSMnet \cite{yoon2018quantitative} and  Gong et al \cite{gong2018qsm}, ASPEN leverages the accurate and computationally efficient forward model to build large training datasets with prescribed characteristics. Networks that rely on COSMOS or STI data are ultimately limited by the accuracy of COSMOS and STI (which are not perfect) and the resolution/features contained within the limited training datasets. 

To ensure data consistency, the ASPEN approach combines the convolutional k-space of field predictions computed from estimated susceptibility maps and measured whole brain field maps in a k-space substitution protocol.   Due to the robustness of the ASPEN neural net, the implemented k-space substitution was not heavily dependent on the size of the substitution region or method of substitution.  Therefore a direct unfiltered substitution matching the TKD threshold of 0.2 was utilized.  This is a substantially different implementation compared to other published k-space substitution approaches in QSM, which have required complex segmentation and combination procedures~\cite{sato2017quantitative}.    

Memory management of the neural network training process has been a challenge for all published QSM neural nets. Though it is possible to train on smaller grids and then infer on larger grids, we found that the parcellated estimation and combination can produce more robust and better QSM results. We have found greater challenges in achieving clean parcel combinations using smaller parcels, due to smaller overlapping regions and increased number of merge regions. Compared with DeepQSM and QSMnet, ASPEN utilizes a larger patch size of 192x192x64 and 128x128x128,
which was found to provide the best QSM estimation while limiting the required GPU resources.

Quantitative accuracy evaluation using the estimated gold-standard dataset from the QSM Challenge showed that the quantitative performance of ASPEN was of the same order, but not substantially better than existing techniques.  This was a curious result, that requires further consideration.   As noted in the QSM Challenge report~\cite{langkammer2018quantitative}, it was surprising how well the CFL2 algorithm quantitatively performed relative to other methods.   The multi-rater large-cohort study performed in the present work adds further confusion to this observation, as the CFL2 approach conclusively was rated to be the worst of all methods in both streaking and resolution performance.    This disparate performance result in the QSM Challenge quantification comparison and basic map quality metrics highlights a risk in over-emphasizing the COMOS or STI gold-standard comparisons.  While a method such as CFL2 may provide better quantification agreement with STI and COSMOS in a variety of global and local metrics in a pristine dataset, it may still not perform at sufficient levels to be of use for generalized research or clinical studies on large study cohorts.       

While the present study has demonstrated that deep-learning based QSM reconstruction can produce super susceptibility estimation compared with commonly utilized methods, there are still some limitations. First, multiple networks must be trained to provide flexibility in prescribed resolution of QSM. For example, lower non-isotropic resolution networks will need to be trained to accommodate data collected within existing susceptibility-weighted clinical acquisition. Future work will investigate the level of tolerance for networks trained on resolutions that are not matched to resolution of acquired data.

\section{CONCLUSIONS}
ASPEN, a volume parcellated deep-learning QSM inversion method has been introduced and evaluated in preliminary fashion. Quantitative and qualitative performance evaluations in this work demonstrate that the ASPEN can provide improved map quality with similar levels of quantitative accuracy compared to commonly utilized QSM methods. 

\section*{Acknowledgements}
The authors would like to acknowledge Dr. Michael McCrea for allowing the use and evaluation of QSM data from a sports concussion study under his direction.  



\bibliographystyle{unsrt}
\renewcommand{\refname}{Bibliography and References Cited }
\bibliography{sample}

\newpage

\begin{table}
\caption{\textbf{\small {Table 1. Numerical measures of QSM reconstruction quality compared to the STI (3,3) ground truth estimate for 2016 QSM Reconstruction Challenge dataset. The last column is the metric achieved by the winning approach for each evaluation test. }}}
\label{qsmChallengeMapsMetrix}
\vspace{0.1in}
\begin{tabular}{cccccc}
\hline
\multicolumn{1}{|c}{} & \multicolumn{1}{|c}{TKD} & \multicolumn{1}{|c}{MEDI} & \multicolumn{1}{|c}{CFL2} & \multicolumn{1}{|c}{ASPEN} & \multicolumn{1}{|c|}{Winning} \\
\hline 
\multicolumn{1}{|c}{RMSE ($\%$)} & \multicolumn{1}{|c}{86.5} & \multicolumn{1}{|c}{126.5} & \multicolumn{1}{|c}{81.2} & \multicolumn{1}{|c}{90.8} & \multicolumn{1}{|c|}{69.0} \\
\multicolumn{1}{|c}{HFEN ($\%$} & \multicolumn{1}{|c}{82.0} & \multicolumn{1}{|c}{121.6} & \multicolumn{1}{|c}{75.5} & \multicolumn{1}{|c}{85.5} & \multicolumn{1}{|c|}{63.5} \\
\multicolumn{1}{|c}{SSIM (0-1)} & \multicolumn{1}{|c}{0.77} & \multicolumn{1}{|c}{0.83} & \multicolumn{1}{|c}{0.81} & \multicolumn{1}{|c}{0.77} & \multicolumn{1}{|c|}{0.94} \\
\multicolumn{1}{|c}{ROI-Error (ppb)} & \multicolumn{1}{|c}{21} & \multicolumn{1}{|c}{22} & \multicolumn{1}{|c}{18} & \multicolumn{1}{|c}{25} & \multicolumn{1}{|c|}{16} \\
\hline
 \end{tabular}
 \begin{tablenotes}
\item RMSE: root mean squared error, HFEN: high-frequency error norm, SSIM: structure similarity index. 
\end{tablenotes}
\end{table}

 \begin{table}
\caption{\textbf{\small {Table 2. Composite results of multi-rater blinded ranking of QSM for streaking artifacts and map sharpness.  Scoring criteria: 1$\rightarrow$worst, 4$\rightarrow$best.   P.A$\rightarrow$ Percentage of cases where all raters had the same score for a given method.  Rater ICC$\rightarrow$ pooled intra-class correlation across raters for all subjects. Cohort statistics for each score are displayed as: mean [std. deviation]. }}}
\label{raterMetrix2}
\vspace{0.1in}
\begin{tabular}{cccccccccc}
\hline
\multicolumn{1}{|c}{} & 
\multicolumn{1}{|c}{TKD} & 
\multicolumn{1}{c}{P.A.} & 
\multicolumn{1}{|c}{CFL2} & 
\multicolumn{1}{c}{P.A.} & 
\multicolumn{1}{|c}{MEDI} & 
\multicolumn{1}{c}{P.A.} & 
\multicolumn{1}{|c}{ASPEN} & 
\multicolumn{1}{c}{P.A.} &
\multicolumn{1}{|c|}{Rater ICC} 
\\
\hline
\multicolumn{1}{|c}{Streaking} & 
\multicolumn{1}{|c}{2.6 [0.6] } & 
\multicolumn{1}{c}{18$\%$} & 
\multicolumn{1}{|c}{1.1 [0.3] } & 
\multicolumn{1}{c}{76$\%$} & 
\multicolumn{1}{|c}{2.3 [0.6]} & 
\multicolumn{1}{c}{14$\%$} & 
\multicolumn{1}{|c}{3.9 [0.1]} & 
\multicolumn{1}{c}{87$\%$} &
\multicolumn{1}{|c|}{0.81 [0.14]} 
\\
\hline
\hline
\multicolumn{1}{|c}{Sharpness} & 
\multicolumn{1}{|c}{3.2 [0.5] } & 
\multicolumn{1}{c}{40$\%$} & 
\multicolumn{1}{|c}{1.1 [0.3] } & 
\multicolumn{1}{c}{82$\%$} & 
\multicolumn{1}{|c}{2.0 [0.4]} & 
\multicolumn{1}{c}{71$\%$} & 
\multicolumn{1}{|c}{3.8 [0.5]} & 
\multicolumn{1}{c}{44$\%$} &
\multicolumn{1}{|c|}{0.86 [0.16]}  
\\
\hline 
 \end{tabular}
\end{table}

\end{document}